\documentclass[%
 aps,
 jcp,
 jmp,%
 amsmath,amssymb,
 reprint,%
]{revtex4-1}

\usepackage{graphicx}
\usepackage{dcolumn}
\usepackage{bm}
\usepackage[mathlines]{lineno}
\usepackage{subfig}
\draft
\begin{document}

\preprint{AIP/123-QED}

\title{Multistability, multiperiodicity, multichaos: in a unified framework}

\author{Feng Liu}
\author{Zhi-Hong Guan}
\email{zhguan@mail.hust.edu.cn(Z.-H. Guan)}
\affiliation{Department of Control Science and Engineering\\
Huazhong University of Science and Technology, Wuhan, 430074, P. R. China}

\date{\today}

\begin{abstract}
In this paper, we present a unified framework of multiple attractors including multistability, multiperiodicity and multichaos. Multichaos, which means that the chaotic solution of a system lies in different disjoint invariant sets with respect to different initial values, is a very interesting and important dynamical behavior, but it is never addressed before to the best of our knowledge. By constructing a multiple logistic map, we show that multistability, multiperiodicity and multiple chaos can exist according to different value of the parameter $p$. In the end, by the derived compact invariant set of the Lorenz system, the multiple Lorenz chaotic attractors are constructed using a sawtooth function.
\end{abstract}
\maketitle
\section{Introduction}
Ever since the establishment of classical control theory by several famous scientists, namely Nyquist, Bode, Harris, Evans, Wienner, Nichols et al in the beginning of 20th century, most of the existing works have studied the mono-stability and mono-periodicity of the dynamical system. However, many physical and biological systems are known in which there are a multitude of coexisting attractors \cite{Ngonghala11,Park99,Maistrenko07,Yi11,Singh08}. Examples include systems from laser physics, chemistry, semiconductor physics, neuroscience, and population dynamics, see reference \cite{Feudel08} and the references therein. Due to its significant applications on neural networks (NNs) with respect to associate memory pattern, recognition and decision making, the topics on multistability and multiperiodicity of NNs began to gain a lot of attention and have been investigated intensively in the past decade
\cite{Cao08,Huang101,Wang10,Zeng06,Zhang09}. Moreover, in a sense, multistability is a necessary property in NNs in order to enable certain applications where mono-stability networks could be computationally restrictive \cite{Hahnloser98}.

Due to the cornucopia of opportunities and rich flexibility provided by the chaotic system, chaos, as a very special dynamical behavior in nonlinear system, has been thoroughly investigated in the recent years\cite{Chen03,Liu09new,Liu09,Liu11,Guan02,Guan06,GuanLiu10,GuanLiu12,Chen0314,Sinha98,Sinha01}. Topics include stabilization, anti-control and synchronization of chaotic system. Although their are a lot of research about chaos, multichaos, which means that the chaotic solution of a system lies in different disjoint invariant sets with respect to different initial values has never been talked and studied. Multistability, multiperiodicity as well as multichaos are different properties of the solutions in disjoint invariant sets, the research on multichaos in systems with disjoint invariant sets has the same significance as the research of chaos in mono-system. Although there should be multichaos solution of a system after the multistability and multiperiodicity solution have been found, nobody has given an exact example to demonstrate the multichaos phenomenon thus far. In this paper, by construction of a multiple logistic map, we can observe the multichaos phenomenon which help us understand the more complicated case in dynamical systems.

\section{A multiple logistic map}
\medskip

\subsection{definitions}
\emph{Definition 1:} Let $S$ be a compact subset of $R^n$. The $S$ is said to be a positive invariant set of the $n$ dimensional continuous system $\dot x = f(x)$ or the discrete map $x(n + 1) = f(x(n))$ if their solution trajectories will not get out once entered $S$.

\emph{Definition 2:} If there exist two or more disjoint positive invariant sets for the $n$ dimensional continuous system $\dot x = f(x)$ or the discrete map $x(n + 1) = f(x(n))$ in $R^n$, and the solution in every disjoint set is chaotic, then this phenomenon is called multichaos and the above systems are called multichaotic system.

\emph{Remark 1:}The multichaotic system is different from multi-wing chaotic system discussed by Simin Yu et al \cite{Yu10}, Xinzhi Liu et al \cite{Liu12}, where the multi-wing chaotic system contains only one positive invariant set.

\subsection{multiple attractors in multiple logistic map}
With the advent of fast computers, the numerical investigations of chaos have increased considerably over the last three decades and by now, a lot is known about chaotic systems. One of the simplest and most transparent systems exhibiting order to chaos transition is the logistic map. In this subsection, we give a unified framework which is constructed by multiple logistic map to demonstrate multistability, multiperiodicity and multichaos according to the different value of the control parameter $p$.

The logistic map is defined by
\begin{equation}\label{1}
    x(n + 1) = f(x(n))=px(n)(1 - x(n)),
\end{equation}
where $x(n)$ a number between zero and one, and represents the ratio of existing population to the maximum possible population at year $n$, and hence $x_0$ represents the initial ratio of population to maximum population (at year 0), $p$ is a positive number, and represents a combined rate for reproduction and starvation. Here in this paper, we assume $p$ is a number between zero and four.
As we all know, the Eq.(\ref{1}) will demonstrate stable fixed points, period-2, period-4, period-8, $ \cdots $ and chaos dynamics according to different $p$. Specifically, when $p$ is between $0$ and $3$, there exists a fixed points; when $p$ is between 3 and approximately 3.57, there are periodic oscillations; when $p$ is larger than approximately 3.57, the Eq.(\ref{1}) will undergo chaotic behavior.

\begin{figure}
\centering
\includegraphics[height=6cm,width=9cm]{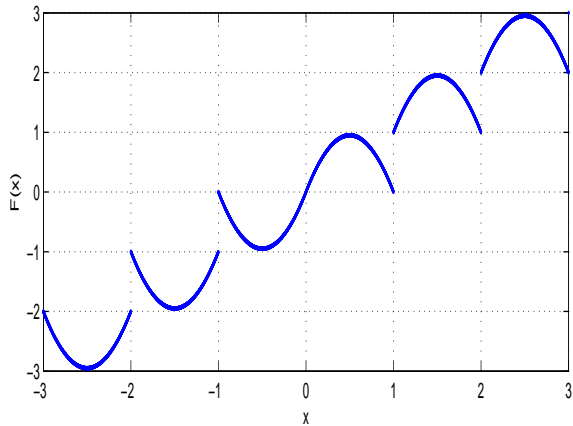}
\caption{value of $F(x)$ as $x$ varies from -3 to 3}
\label{functionf}
\end{figure}

Next, a multiple logistic map $F(x)$ is redefined by connecting $f(x)$ and extending its domain infinitely instead of only in the invariant set $(0,1)$. Below is the definition of $F(x)$:
Let $Z$ be the set of all integers, for any $k \in Z$, when $k \leq x<k+1$
\begin{equation}\label{2}
\left\{ \begin{array}{l}
F(x) = p(x - k)(k + 1 - x) + k,\;\;\quad \quad \,when\;k > 0\\
F(x) =  - p(x - k)(k + 1 - x) + k + 1,\;when\;k < 0
\end{array} \right..
\end{equation}
When $x \in [ - 3,3]$, let $p=3.8$, the figure of $F(x)$ is given in Fig.\ref{functionf}.

For any $x_0 \in (k,k+1)$, the solution of $x(n+1)=F(x(n))$ lies within $(k,k+1)$ beginning the the initial $x_0$.
According to \emph{definition 1}, it is evident that for any $k \in Z$, the set $(k,k+1)$ is positive invariant set of $F(x)$. The proof is trivial, thus it is omitted here. It is also very evident that there are infinite positive invariant sets because there are infinite integers.

\begin{figure}
\centering
\includegraphics[height=6cm,width=9cm]{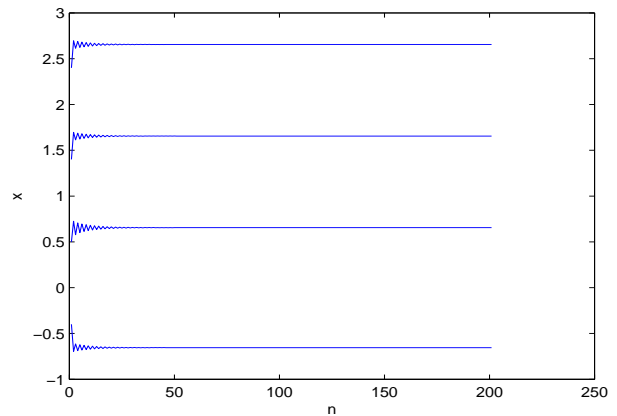}
   \caption{multistability when $p=2.9$}
      \label{ms}
\end{figure}

\begin{figure}
\centering
\includegraphics[height=6cm,width=9cm]{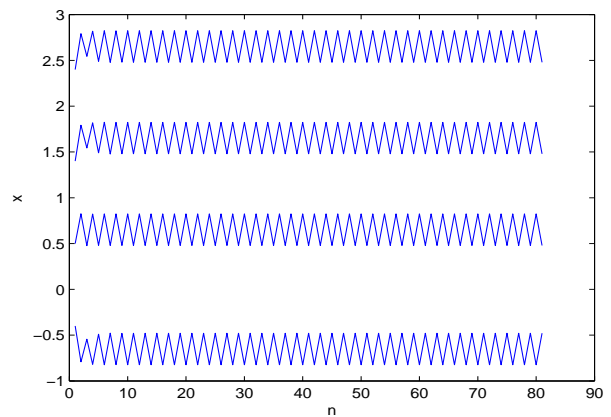}
   \caption{multiple period-2 when $p=3.3$}
      \label{mp2}
\end{figure}

\begin{figure}
\centering
\includegraphics[height=6cm,width=9cm]{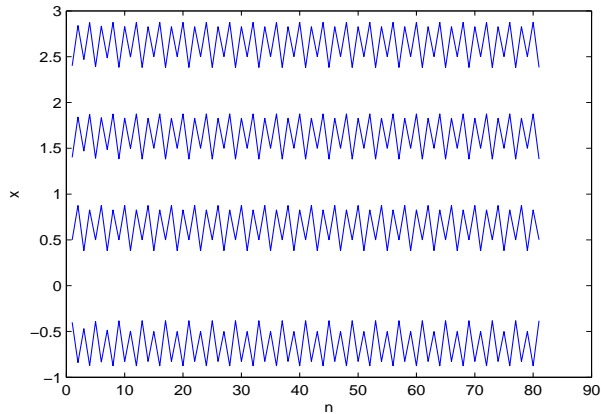}
   \caption{multiple period-4 when $p=3.5$}
      \label{mp4}
\end{figure}

\begin{figure}
\centering
\includegraphics[height=6cm,width=9cm]{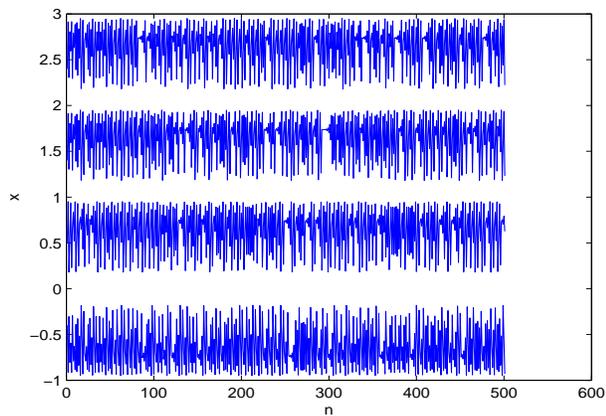}
   \caption{multichaos when $p=3.8$}
      \label{mc}
\end{figure}

Next, we demonstrate that multistability, multiperiodicity and multichaos can exist in the discrete map Eq.\ref{2} using simulation examples. In Fig.\ref{ms}-\ref{mc}, four different initial values which lie in the four positive invariant sets $(-1,0)$, $(0,1)$, $(1,2)$ and $(2,3)$ respectively are chosen for each $p$ in the four figures. The initial values are the -0.4, 0.5, 1.4, 2.4. Fig.\ref{ms} demonstrates multistability when $p=2.9$, with each solution orbit residing on a fixed point which lies in the corresponding positive invariant sets. Fig.\ref{mp2} and Fig.\ref{mp4} shows multiple period-2 and period-4 solution with $p=3.3$ and $p=3.5$ respectively. Fig.\ref{mc} presents multichaos phenomenon when $p=3.8$.

\section{Construction of multichaos based on Lorenz system}
\begin{figure}
\centering
\includegraphics[height=6cm,width=9cm]{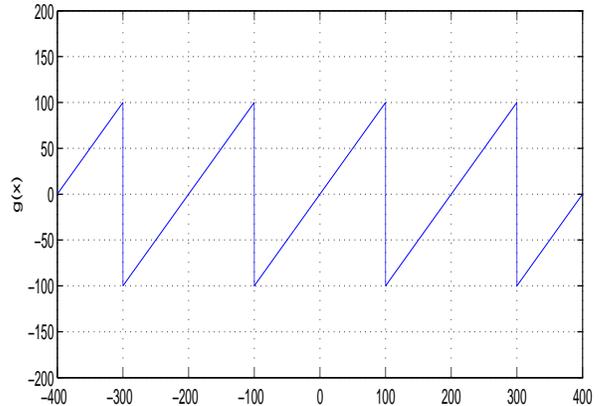}
\caption{value of $g(x)$ as $x$ varies from -400 to 400}
\label{gx}
\end{figure}

In the last section, we give an example to show the existence of multichaos in an one-dimensional map. In this section, a continuous three-dimensional system that exhibits multichaotic behavior is constructed based on Lorenz system. By defining a sawtooth function, the phase space is divided into infinite squares. In each square, there is a positive invariance set in which Lorenz attractors are defined.
The well known Lorenz system \cite{Lorenz63} is given below:
\begin{equation}\label{31}
\begin{array}{l}
\dot x = \sigma (y - x),\\
\dot y = x(\rho  - z) - y,\\
\dot z = xy - \beta z.
\end{array}
\end{equation}
When $\sigma = 10$, $\beta=8/3$ and $\rho=28$, the system \ref{31} exhibits chaotic behavior. According to Ref.\cite{Yu09}, Lorenz system has only one positive invariance set. Let $A$ be the positive invariance set, then the set $A$ is given below by calculating the result derived by Ref.\cite{Yu09}, which is  $A=\{x,y,z|x^2+y^2+(z-38)^2 \leq 1540.3\}$. Let $B=\{x,y,z|-100<x,y,z<100\}$. Apparently, $A$ is the subset of $B$. When the initial value $(x(0),y(0),z(0))$ belongs to $A$, the trajectory of \ref{31} will not get out according to the definition of positive invariant set. In order to generate Lorenz-like multichaotic attractors, a sawtooth function and a novel multichaotic lorenz system is defined. The sawtooth function $g(x)$ is defined as below:
\begin{equation}\label{32}
g(x)=mod(x+100,200)-100.
\end{equation}
When $-400<x<400$, $g(x)$ is depicted in Fig.\ref{gx}.

\begin{figure}
\centering
\includegraphics[height=6cm,width=9cm]{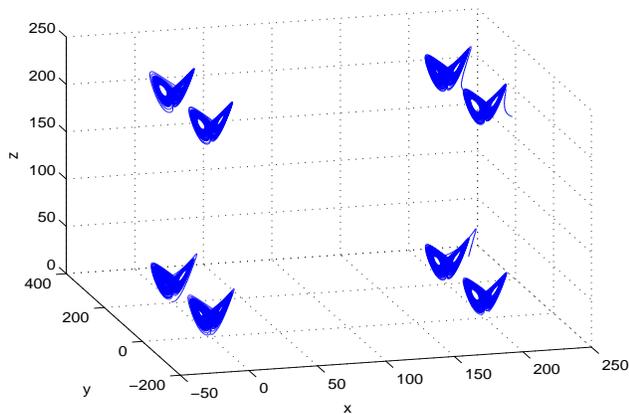}
\caption{Coexistence of eight chaotic attractors}
\label{mlc}
\end{figure}

For simplicity, denote $X=g(x)$, also, $Y$ and $Z$ are defined likewise. The multichaotic system based on Lorenz system is given as below:
\begin{equation}\label{33}
\begin{array}{l}
\dot x = \sigma (Y - X),\\
\dot y = X(\rho  - Z) - Y,\\
\dot z = XY - \beta Z.
\end{array}
\end{equation}
When the initial value lies in the compact set $A$, the solution of \ref{33} will be exactly like \ref{31} and thus exhibiting a typical double wing strange attractor. Denote $C=\{(x,y,z)|(x-200k)^2+(y-200m)^2+(z-38-200n)^2 \leq 1540.3, \forall k,m,n \in Z \}$. It is obvious that $C$ is composed of infinite disjoint compact subsets. For an initial point $(x_0,y_0,z_0)$ starting from any of the disjoint sets of $C$, there is a transformation satisfying $(x_{0}-200k,y_0-200m,z_0-38-200n) \in A$. Let $P=x-200k$,$Q=y-200m$,$R=z-38-200n$, then the equation \ref{33} with the initial point lying in one of the subsets of $C$ will be transformed into the following equation:
\begin{displaymath}
\begin{array}{l}
\dot P = \sigma (Q - P),\\
\dot Q = P(\rho  - R) - Q,\\
\dot R = PQ - \beta R,
\end{array}
\end{displaymath}
which is the same with the case of Eq.\ref{31} with a initial point starting from $A$. It is very easy to prove that any orbit starting from a initial point lying in $C$ will not get out of a positive invariant set determined by the specific value of $k,m,n$.
A simulation result is given to verify the coexistence of multichaotic attractors in Fig.\ref{mlc}. There are eight chaotic attractors in Fig.\ref{mlc}, with each starting from different positive invariant subsets of $C$.

\section{Conclusions}
In this paper, we give the definition of multichaos which is never defined before. By constructing a multiple logistic map, we show that the map can exhibit multistability, multiperiodicity and multichaos according to different $p$. Then the continuous multiple Lorenz chaotic attractors are constructed using a sawtooth function. By calculating the positive invariant sets, we show that initial points starting in the positive invariant sets will not get out and will exhibit chaotic dynamical behavior in each disjoint positive invariant set.
\medskip

\begin{acknowledgments}
This work was supported in part by the National Natural Science Foundation of China under Grant 60973012, 61073025, 61073026, 61074124, 61170031 and the Graduate Student Innovation Fund of Huazhong University of Science and Technology under Grant 0109184979.
\end{acknowledgments}

\end{document}